\documentclass[aps,prd,preprintnumbers,superscriptaddress,nofootinbib,notitlepage]{revtex4-2}
\usepackage[pdftex]{graphicx}
\usepackage{bm,latexsym,amsmath,amssymb,amsfonts,mathrsfs}
\usepackage{mathtools}
\usepackage{multirow}
\usepackage{color}
\allowdisplaybreaks[1]
\usepackage[pdftex,colorlinks=true,linkcolor=blue,citecolor=cyan]{hyperref}

\begin{document}
\title{Higher-dimensional quantum-corrected Oppenheimer-Snyder model with a cosmological constant}

\author{Shudi Jiang}
\email[Email: ]{sdjiang03@gmail.com}
\affiliation{School of Physics and Astronomy, University of Edinburgh, Edinburgh, United Kingdom
}

\author{Jianhui Lin}
\email[Email: ]{202510188472@mail.scut.edu.cn}
\affiliation{School of Physics and Optoelectronics, South China University of Technology, Guangzhou 510641, China
}

\author{Xiangdong Zhang}
\email[Email: ]{scxdzhang@scut.edu.cn}
\affiliation{School of Physics and Optoelectronics, South China University of Technology, Guangzhou 510641, China
}

\begin{abstract}
We extended the higher-dimensional quantum Oppenheimer-Snyder model to the case with a cosmological constant. 
For AdS case, we discuss its thermodynamic properties in extend phase space formalism and make comparison with classical black holes. For quantum-corrected small black holes in AdS spacetime, the temperature no longer diverges but tends to zero. Additionally, the heat capacity exhibits characteristic behavior indicative of an extra phase transition induced by quantum corrections, highlighting the profound impact of quantum effects on black hole thermodynamics.
\end{abstract}

\preprint{RUP-23-14}
\maketitle


\section{Introduction}
Since the 20th century, quantum theory and Einstein's general theory of relativity—the two major branches of modern physics—have each achieved success at the extreme microscopic and extreme gravitational scales, respectively. However, the unification of these two theories: the construction of a self-consistent quantum gravity theory—remains as one of the central unresolved challenges in theoretical physics to this day. Although general relativity perfectly describes classical gravitational phenomena ranging from stellar evolution to large-scale cosmic structures, it fundamentally fails when predicting spacetime singularities such as the centers of black holes and the Big Bang singularity \cite{Carballo-Rubio:2018jzw,Vasconcellos:2019kan}. This indicates that at the Planck scale, quantum gravitational effects must be introduced to address the limitations of classical theory.

Loop quantum gravity (LQG) is one of the most promising approaches to quantum gravity. It offers a different view of spacetime at the quantum level and brings new possibilities for resolving singularities \cite{Thiemann:2001gmi,Ashtekar:2004eh,Han:2005km}. In recent years, remarkable progress has been made in this field, particularly in studies of the quantum nature of black holes \cite{Ashtekar:1997yu,Ashtekar:2005qt,Rovelli:1996dv,Zhang:2023yps,Lin:2024beb,Du:2024ujg,Du:2025kcx}.

Within classical general relativity, the gravitational collapse of massive stars is often studied using the Oppenheimer-Snyder (OS) model \cite{Oppenheimer:1939ue}. This idealized description assumes both spherical symmetry and homogeneous collapse evolution. Even with its simplified setup, the OS model manages to reproduce the essential physical behavior of collapsing stellar objects. One significant prediction arising from this theory is a minimum mass for black hole formation, known as the Tolman–Oppenheimer–Volkoff (TOV) limit \cite{Tolman:1939jz}, which is roughly three solar masses. In recent investigations, a nonsingular four-dimensional black hole has been obtained by coupling the OS collapse scenario with LQG model \cite{Lewandowski:2022zce}. The corresponding spacetime experiences a transition from black hole to white hole \cite{Han:2023wxg}. This model has quickly been generalized to the higher dimensional case \cite{Shi:2024vki}. 

Black hole thermodynamics has emerged as a profound and active field at the interface of general relativity, quantum theory, and statistical mechanics. In particular, the discovery of Hawking radiation \cite{Hawking:1974rv,Hawking:1975vcx} revealed that black holes are not purely classical geometric ensembles, but rather thermodynamic systems endowed with well-defined temperature and evaporation behavior. Correspondingly, the four thermodynamic laws of a black hole \cite{Bardeen:1973gs} were well established, exhibiting striking analogies with the laws of ordinary thermodynamics. Among these, the Bekenstein-Hawking entropy plays a central role, providing a profound link between black hole horizon geometry and microscopic quantum gravitational degrees of freedom \cite{Bekenstein:1973ur, Rovelli:1996dv,Strominger:1996sh}. 
	
Current cosmological observations on large scales indicate that the expansion of the Universe is accelerated, suggesting the need to introduce dark energy in cosmological models \cite{SupernovaSearchTeam:1998fmf,Weinberg:1988cp,Peebles:2002gy}. Among various explanations, the cosmological constant remains the most widely accepted one. A positive cosmological constant, corresponding to the de Sitter (dS) spacetime, is consistent with observational data, while a negative cosmological constant is closely related to the AdS/CFT correspondence \cite{Maldacena:1997re,Witten:1998qj}, which states that black holes in anti-de Sitter (AdS) spacetime exhibit remarkably rich thermodynamic behaviors, particularly in various phase transitions and critical phenomena \cite{Kubiznak:2012wp,Cai:2013qga,Wei:2014hba,Wei:2017icx}. One of the representative phase transitions, the Hawking–Page phase transition serves as a typical example \cite{Hawking:1982dh, Li:2020khm}, describing a first-order phase transition between thermal AdS spacetime and large-mass AdS black holes, thus providing important insights into the thermodynamic stability along with the phase structure of black holes. Inspired by the above motivations, in this work, we focus on extending the higher dimensional OS model to incorporate a negative cosmological constant (anti-de Sitter cases) and systematically investigate its thermodynamic properties.  
	
This paper is organized as follows. First, we obtain the new metric by combining the Friedmann equation of loop quantum cosmology in higher dimensions with the corresponding quantum Oppenheimer-Snyder (qOS) model in Sec.~\ref{Metric with cosmological constant}. 
After that, we examine the thermodynamic properties of the model in Sec.~\ref{Thermaldynamic properties} and compute the critical exponents in Sec.~\ref{Critical index}. Finally, we summarize and discuss our main results in Sec.~\ref{Conclusion}. Throughout the paper, we work in the geometric units in which $G = c = \hbar = 1$.

\section{Metric with Cosmological Constant}\label{Metric with cosmological constant}

Classical spherically symmetric gravitational collapse is described by the Oppenheimer-Snyder model, which characterizes the relativistic collapse process of a homogeneous, spherically symmetric, non-rotating, pressureless dust ball under its own gravity. The full \(d+1\) dimensional spacetime is divided into two regions by the dust ball surface: the interior fluid region is described by the homogeneous and isotropic Friedmann-Robertson-Walker (FRW) metric, while the exterior vacuum region is characterized by the static spherically symmetric metric, and the two regions are self-consistently connected via the Darmois-Israel junction conditions \cite{israel1966singular}. The metric of the dust ball interior can thus be written as
\begin{equation}
\mathrm{d}s_{\rm in}^{2} = -\mathrm{d}T^{2} + a^{2}(T)\left[\mathrm{d}R^{2} + R^{2} d\Omega^{2}\right],
\label{FRW}
\end{equation}
where the scale factor $a(T)$ encodes the full dynamical evolution of spacetime, as determined by the Friedmann equations for the FRW ansatz. Here, $d\Omega^2$ is the line element of the $(d-1)$-dimensional unit sphere, given by
\begin{align}
    \mathrm{d}\Omega^{2} &= \mathrm{d}\theta_1^{2} + \sin^2\theta_1 \mathrm{d}\theta_2^{2} + \dots \nonumber\\
    &\quad + \sin^2\theta_1 \sin^2\theta_2 \dots \sin^2\theta_{d-2} \mathrm{d}\theta_{d-1}^{2}.
\end{align}
The coordinate $\theta_2$ ranges over $[0, 2\pi)$, while all other angular coordinates are $[0, \pi]$. We define the physical radius of the dust ball as $\tilde{R} = a(T) R_0$, where $R_0$ is the radial coordinate of the dust ball surface. The hypersurface $\Sigma$ separating the interior and exterior spacetimes is defined by the dust ball surface at $R=R_0$. The induced metric on $\Sigma$ from the interior FRW metric (\ref{FRW}) is then given by
\begin{equation}
\left. \mathrm{d}s_{\rm in}^2 \right|_{\Sigma} = -\mathrm{d}T^2 + a^2(T) R_0^2 \mathrm{d}\Omega^2.
\label{hypers1}
\end{equation}
Meanwhile, the metric of the exterior spacetime can be written as
\begin{equation}
\mathrm{d}s_{\rm out}^{2} = -f(r)dt^{2} + g(r)^{-1}\mathrm{d}r^{2} + r^2 \mathrm{d}\Omega^2,
\label{ds_out}
\end{equation}
and the induced metric on the same hypersurface $\Sigma$ (obtained from the exterior spacetime) is given by
\begin{equation}
\left. \mathrm{d}s_{\rm out}^2 \right|_{\Sigma} = -\left(f \dot{t}^2 - g^{-1} \dot{r}^2\right) \mathrm{d}T^2 + r^2(T) \mathrm{d}\Omega^2,
\label{ind_out}
\end{equation}
where we parametrize both $r$ and $t$ as functions of the comoving time $T$ on $\Sigma$ and their derivatives with respect to $T$ as well.

The next step is to match the interior and exterior spacetimes using the Darmois-Israel junction conditions to obtain a complete and self-consistent spacetime. This implies that both the metric and the extrinsic curvature must be continuous across the hypersurface $\Sigma$,
\begin{align}
\left. \mathrm{d}s_{\rm in}^2 \right|_{\Sigma} &= \left. \mathrm{d}s_{\rm out}^2 \right|_{\Sigma}, \\
K^{\rm in}_{\alpha \beta}&=K^{\rm out}_{\alpha \beta}.
\end{align}
The continuity of the metric directly yields two relations:
\begin{equation}\label{gcontinue}
1 = f\dot{t}^2 - g^{-1}\dot{r}^2, \quad a(T)R_0 = r(T).
\end{equation} We then impose the continuity of the extrinsic curvature. From the interior spacetime, the non-zero components of the extrinsic curvature are given by:
\begin{equation}
    \frac{K_{\theta_1\theta_1}^{in}}{\sin^2\theta_1} = \frac{K_{\theta_2\theta_2}^{in}}{\sin^2\theta_1\sin^2\theta_2} = \cdots = a(T)R_0.
\end{equation}
For the exterior static spacetime, the corresponding components of the extrinsic curvature are:
\begin{equation}
    \frac{K_{\theta_1\theta_1}^{out}}{\sin^2\theta_1} = \frac{K_{\theta_2\theta_2}^{out}}{\sin^2\theta_1\sin^2\theta_2} = \cdots = rf\dot{t}\sqrt{f^{-1}g}.
\end{equation}
Equating these two expressions, we obtain the final equation:
\begin{equation}\label{kcontinue}
a(T)R_0 = rf\dot{t}\sqrt{f^{-1}g}.
\end{equation}
A more detailed derivation of these relations can be found in \cite{Shi:2024vki,Ou:2025bbv}.

The above expressions can be further simplified. Since the exterior spacetime is static, it admits a timelike Killing vector field $\xi^\alpha = (\partial/\partial t)^\alpha$. Furthermore, as the dust ball surface is in free fall, the worldline of it follows a geodesic. Consequently, the contraction of the Killing vector with the geodesic tangent vector $u^\alpha = (\partial/\partial T)^\alpha$ produces a conserved quantity $F$, defined as
\begin{equation}
    \xi_\alpha u^\alpha = -f \dot{t} \equiv -F. \label{simplyft}
\end{equation}
With these three Eqs. (\ref{gcontinue}, \ref{kcontinue}, \ref{simplyft}) in hand, we immediately arrive at
\begin{align}
f = F^2 g, \quad g = 1 - \dot{r}^2.
\end{align}
Recalling the Friedmann equations, we find the relation
\begin{align}
    H^2 = \left( \frac{\dot{a}}{a} \right)^2 = \left( \frac{\dot{r}}{r} \right)^2,
    \label{eq:friedmann_hubble}
\end{align}
where the second equality follows from Eq.~\eqref{gcontinue}.
With this result in hand, we can explicitly express the exterior spacetime metric in terms of the Hubble parameter as
\begin{equation}
    \mathrm{d}s^2_{\mathrm{out}} = -\left(1 - H^2 r^2\right) \mathrm{d}t^2 + \left(1 - H^2 r^2\right)^{-1} \mathrm{d}r^2 + r^2 \mathrm{d}\Omega^2,
    \label{metric_out}
\end{equation}
the constant $F$ has been absorbed into a redefinition of the time coordinate $t$.

The quantum-corrected Friedmann equation for $(d+1)$ dimension spacetime \cite{Lewandowski:2022zce} is given by
\begin{equation}
H^{2}=\frac{2\kappa}{d(d-1)} \rho_{T}(1-\frac{\rho_{T}}{\rho_{c}} ),
\label{FE}
\end{equation}
where
\begin{align}
\rho_{T} = \rho+\rho_{\Lambda}, \quad \rho_{\Lambda}=\frac{\Lambda}{\kappa}, \quad \kappa = 8\pi G,\label{rhoT}
\end{align}
and $\rho_{\Lambda }$ is an additional density of matter affected by the cosmological constant $\Lambda$. We denote the critical density in the \((d+1)\) spacetime as \cite{Zhang:2015bxa}
\begin{equation}
\rho_{c}=\frac{d(d-1)}{2\kappa\gamma ^{2}\Delta ^{\frac{2}{d-1} }},
\label{rhoc}
\end{equation}
with $\gamma$ being Immirzi parameter and $\Delta$ area gap \cite{Ashtekar:2008zu}. In the classical limit where the area gap vanishes, i.e., $\Delta \to 0$, it is evident from Eq. (\ref{rhoc}) that the critical density $\rho_c$ diverges to infinity. Consequently, the quantum correction term $\rho_T / \rho_c$ in Eq. (\ref{FE}) becomes negligible, and the quantum-corrected Friedmann equation reduces to its classical counterpart. 

The total matter density \(\rho_T\) in Eq. (\ref{FE}) can also be given by the ratio of mass to volume of the dust ball. If we assume that the dust ball acts as a \(d\)-dimensional sphere with radius $\tilde{R}$, the expression of density becomes
\begin{equation}
\rho_T = \frac{M}{V}= \frac{Md(d-1)\mu}{8\pi\tilde{R}^{d} }  ,
\label{rho}
\end{equation}
with d-dimensional volume
\begin{equation}
V= \frac{8\pi\tilde{R}^{d}}{d(d-1)\mu} ,
\label{Vmatter}
\end{equation}
and
\begin{equation}
\mu =\frac{8\pi}{(d-1)\Omega } , \quad \Omega=\frac{d\pi ^{\frac{d}{2} }}{\Gamma (\frac{d}{2} +1)} .
\label{paraV}
\end{equation}
If we combine Eqs.~\eqref{FE}--\eqref{paraV} and define a correction factor $\alpha\equiv \gamma ^{2}\Delta ^{\frac{2}{d-1} }$, the Friedmann equation can then be written as
\begin{equation}
H^{2} = \frac{2 G M \mu}{r^{d-2}} + \left(\frac{2 G M \mu}{r^{d-1}}\right)^2 \alpha
-\frac{2\left(r^d - 4 G M \mu \alpha\right)}{d(d-1) r^{d-2}} \Lambda + \left(\frac{2 r \Lambda}{d(d-1)}\right)^2 \alpha.
\label{FE2}
\end{equation}
As a result, the metric \eqref{metric_out} becomes
{
\begin{align}
\mathrm{d}s_{\rm out}^{2} = & -\left[1 - \frac{2 G M \mu}{r^{d-2}} + \left(\frac{2 G M \mu}{r^{d-1}}\right)^2 \alpha \right. 
\left. -\frac{2\left(r^d - 4 G M \mu \alpha\right)}{d(d-1) r^{d-2}} \Lambda + \left(\frac{2 r \Lambda}{d(d-1)}\right)^2 \alpha \right]\mathrm{d}t^{2} \nonumber\\
& +\left[1 - \frac{2 G M \mu}{r^{d-2}} + \left(\frac{2 G M \mu}{r^{d-1}}\right)^2 \alpha \right. \left. -\frac{2\left(r^d - 4 G M \mu \alpha\right)}{d(d-1) r^{d-2}} \Lambda + \left(\frac{2 r \Lambda}{d(d-1)}\right)^2 \alpha \right]^{-1}\mathrm{d}r^{2} \nonumber\\
& + r^{2}\mathrm{d}\Omega.
\label{dsfinal}
\end{align}
}

\section{Thermodynamic Properties}\label{Thermaldynamic properties}
\subsection{Temperature}
The event horizon $r_h$ is determined by the condition $f(r_h) = 0$. This relation allows the black hole mass $M$ to be expressed in terms of the horizon radius $r_h$. We require the solution $M(r_h)$ to remain well-behaved in the limit where both the correction factor $\alpha$ and the cosmological constant $\Lambda$ vanish. The resulting expressions in various dimensions are summarized in Table~\ref{star_mass}.
\begin{table}[htbp]
	\centering
	\caption{Dust ball mass expressions in different dimensions with $\Lambda$.}
	\label{star_mass}
	\renewcommand{\arraystretch}{2.0} 
	\setlength{\tabcolsep}{20pt} 
	\normalsize
	\begin{tabular}{cc}
		\hline\hline
		\textbf{(d+1)-dimension} & \textbf{Mass $M(r_h)$} \\
		\hline
		4 & $-\frac{1}{6}r_{h}\bigl(-3+r_{h}^{2}\Lambda\bigr) + \frac{\alpha}{2r_{h}} + \mathcal{O}(\alpha^2)$ \\
		5 & $-\frac{1}{16}\pi r_{h}^{2}\bigl(-6+r_{h}^{2}\Lambda\bigr) + \frac{3\pi\alpha}{8} + \mathcal{O}(\alpha^2)$ \\
		6 & $-\frac{1}{15}\pi r_{h}^{3}\bigl(-10+r_{h}^{2}\Lambda\bigr) + \frac{2\pi r_{h}\alpha}{3} + \mathcal{O}(\alpha^2)$ \\
		7 & $-\frac{1}{48}\pi^{2}r_{h}^{4}\bigl(-15+r_{h}^{2}\Lambda\bigr) + \frac{5}{16}\pi^{2}r_{h}^{2}\alpha + \mathcal{O}(\alpha^2)$ \\
		\hline\hline
	\end{tabular}
\end{table}

The Hawking temperature of a black hole is related to the surface gravity $k_{h}$ at the event horizon $r = r_h$ \cite{Mirekhtiary:2014qnj}. For a given metric function $f(r)$, the surface gravity is defined as
\begin{equation}
k_{h} = \left. \frac{f'(r)}{2} \right|_{r=r_{h}},
\label{eq:surface_gravity}
\end{equation}
Consequently, the Hawking temperature can be expressed with \(k_{h}\) 
\begin{equation}
T = \frac{k_{h}}{2\pi}.
\label{eq:Hawking_temperature}
\end{equation}
We would like to compare the quantum-corrected temperature with the classical one, i.e. Schwarzschild temperature. The latter can be obtained by the same process in the Hawking temperature calculation but takes the value of $\alpha$ to be zero.

\begin{figure}[htbp]
	\centering
	\includegraphics[width=0.5\linewidth]{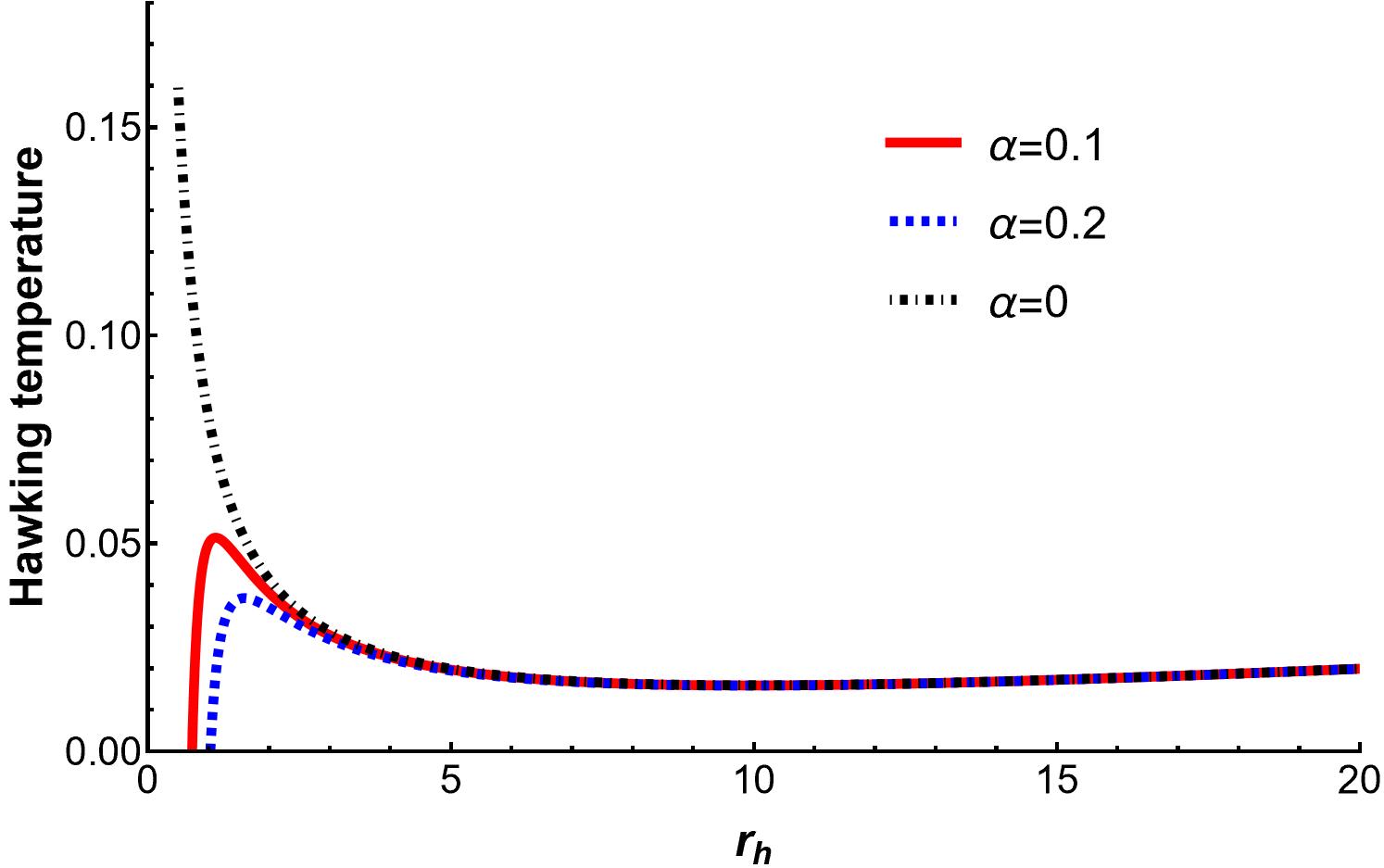}
	\caption{Hawking temperature for the quantum-corrected and Schwarzschild black hole in 5-dimensional spacetime with $\Lambda$=-0.01.}
	\label{HT5D}
\end{figure}

\begin{figure}[htbp]
	\centering
	\includegraphics[width=0.5\linewidth]{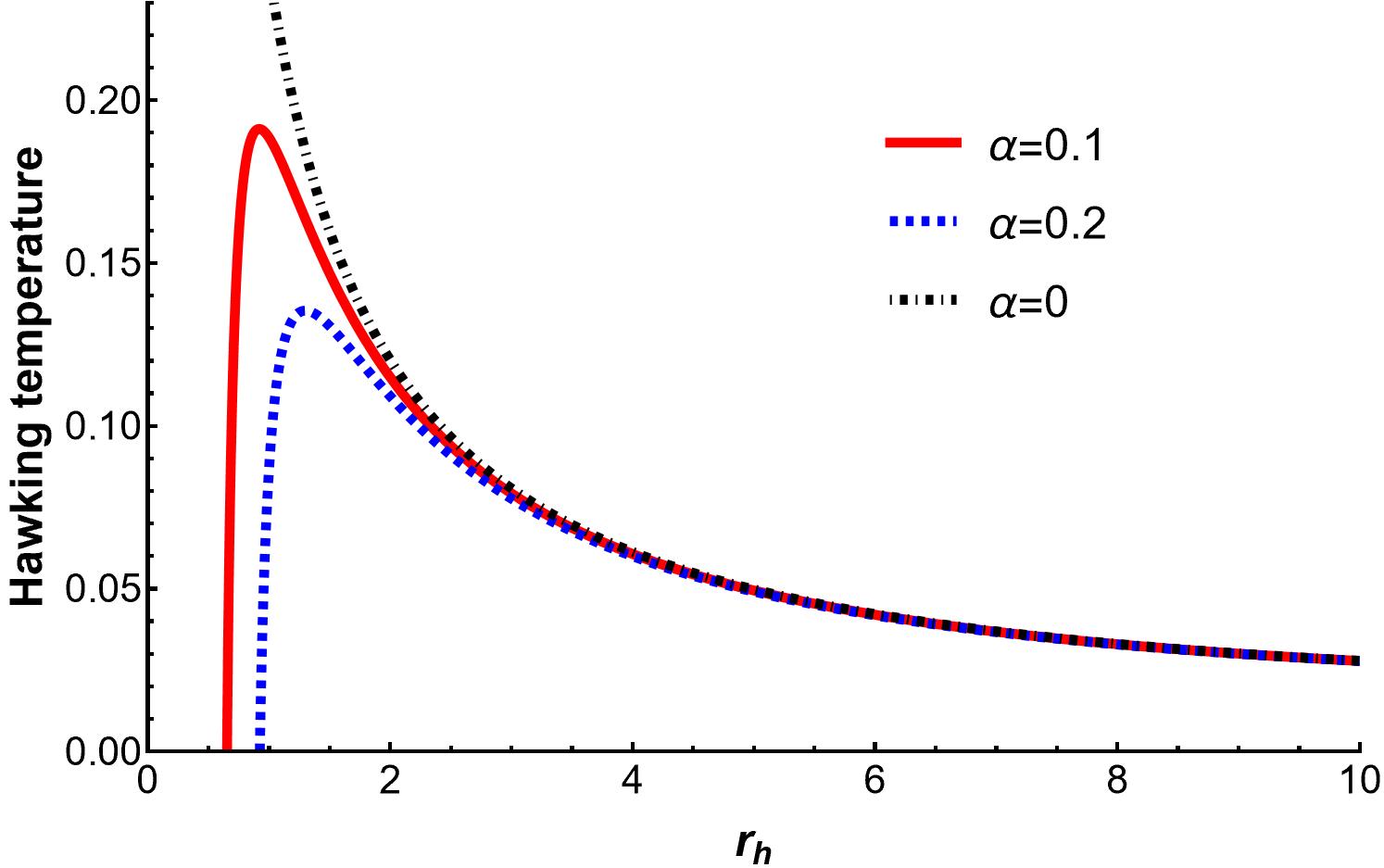}
	\caption{Hawking temperature for the quantum-corrected and Schwarzschild black hole in 6-dimensional spacetime with $\Lambda$=-0.01.}
	\label{HT6D}
\end{figure}

As shown in Figs.~\ref{HT5D} and ~\ref{HT6D}, we find that for small-radius cases, the temperature in the classical model diverges as \(r_h \to0\), whereas the temperature with quantum corrections increases rapidly from zero to a peak. This non-monotonic temperature profile at small $r_h$ differs fundamentally from the classical prediction and could potentially indicate a key modification to the evaporation process. Specifically, the vanishing temperature at a finite scale suggests that the evaporation of small black holes may not proceed to completion. Instead, the process could be halted at this characteristic scale, leaving a long-lived, non-evaporating remnant-like configuration. Moreover, the observed phenomenon shows that the radius of the black hole remnant increases with the parameter \(\alpha\), which points to a close relationship between the quantum correction effects and the black hole remnant. As $r_{h}$ continues to increase, the temperature with quantum corrections asymptotically approaches that of the classical result. Furthermore, we observed that, beyond a certain radius, the temperature begins to rise slowly, as illustrated in Fig.~\ref{HT5D}. This qualitative behavior remains consistent in higher dimensions. In Fig.~\ref{HT6D}, the peak values of the quantum-corrected temperature increase with dimension, and it should be noted that as the dimension increases, the subsequent minimum of the temperature shifts to a larger radius.


\subsection{Entropy}\label{Entropy}
In the framework of extended phase space thermodynamics, the first law of thermodynamics is $dM=TdS+VdP$, and the pressure connected to the cosmological constant $\Lambda$ is defined by \cite{dolan2011cosmological}
\begin{align}
    P=\frac{-\Lambda}{8\pi},
\end{align}
which is independent of the spacetime dimension. If we assume that the black hole satisfies standard thermodynamic relations, the entropy becomes \cite{panah2025super}
\begin{equation}
S=\int T^{-1}dM=\int_{r_{0}}^{r_{h}} T^{-1}(\frac{\partial M}{\partial r} )dr.
\label{entropy}
\end{equation}
The lower integration limit, $r_0$, corresponds to the radius at which the entropy vanishes. Physically, we require the entropy to be zero in the zero-temperature limit. Thus, the value of $r_0$ is determined by the condition $T(r_0) = 0$. To examine whether the entropy will be affected by the additional cosmological constant, we list the entropy expressions for different dimensions in Tab.~\ref{entropy_dimensions}. All results are expanded in the limit of small $\alpha$.

\begin{table}[htbp]
	\centering
	\caption{Black hole entropy expressions in different dimensions with quantum correction $\alpha$.}
	\label{entropy_dimensions}
	\renewcommand{\arraystretch}{2.0} 
	\setlength{\tabcolsep}{20pt} 
	\normalsize
	\begin{tabular}{ccc}
		\hline\hline
		\textbf{(d+1)-dimension} & \textbf{Entropy} & \textbf{S/A} \\
		\hline
		5 & $\frac{\pi^{2}r_{h}^{3}}{2} + 3\pi^{2}r_{h}\alpha$ & $\frac{1}{4} + \frac{3\alpha}{2r_h^2}$ \\
		6 & $\frac{2\pi^{2}r_{h}^{4}}{3} + \frac{8}{3}\pi^{2}r_{h}^{2}\alpha$ & $\frac{1}{4} + \frac{\alpha}{r_h^2}$ \\
		7 & $\frac{\pi^{3}r_{h}^{5}}{4} + \frac{5}{6}\pi^{3}r_{h}^{3}\alpha$ & $\frac{1}{4} + \frac{5\alpha}{6r_h^2}$ \\
		\hline\hline
	\end{tabular}
\end{table}
Regarding the analytical expressions for the entropy, the cosmological constant $\Lambda$ cancels out during the derivation in any dimension. Consequently, the entropy is independent of $\Lambda$, and the terms involving $\alpha$ originate exclusively from quantum corrections. 
\begin{figure}[htbp]
	\centering
	\includegraphics[width=0.5\linewidth]{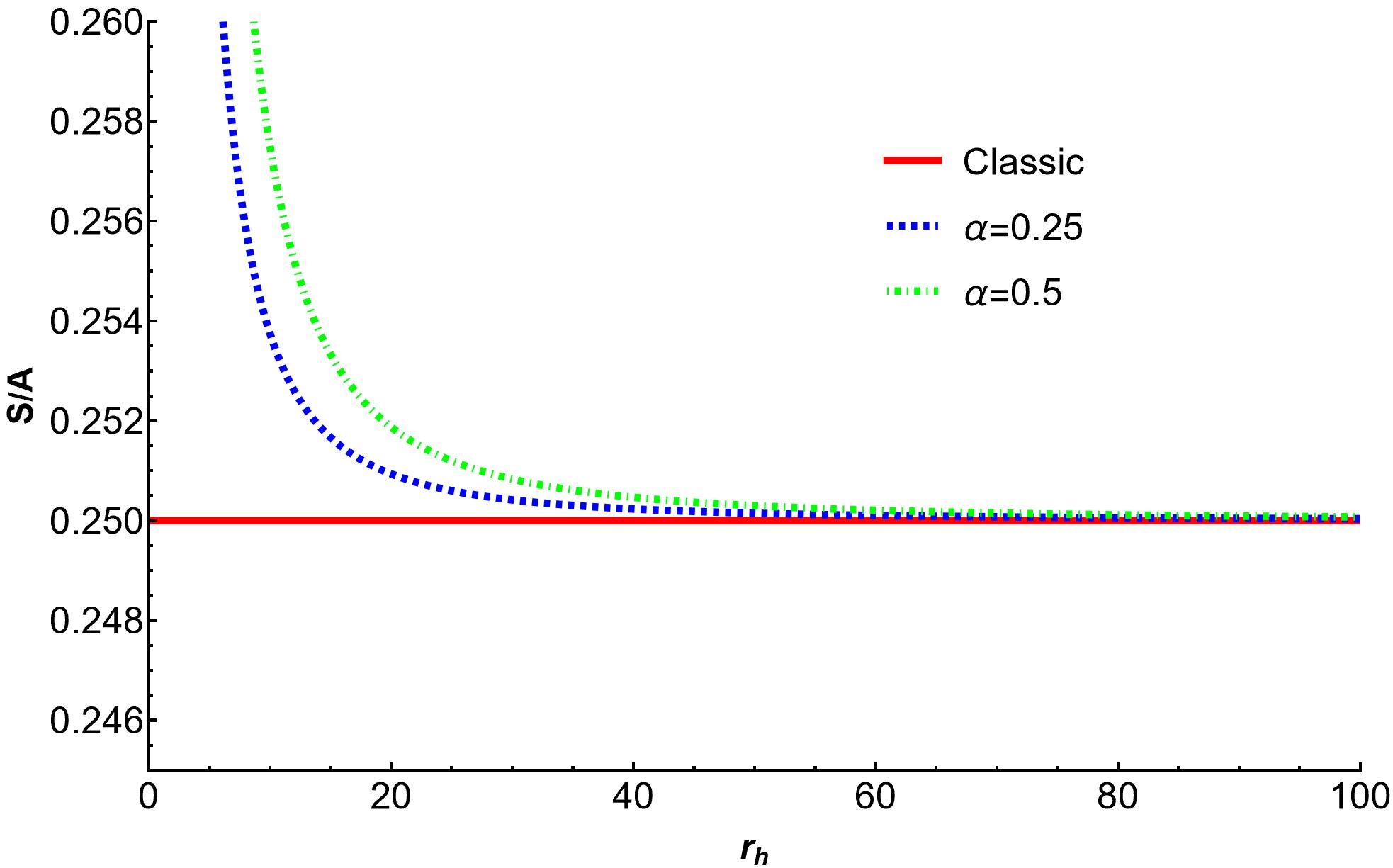}
	\caption{Entropy of quantum-corrected and Schwarzschild (classical) black hole in 5-dimensional space.}
	\label{S5}
\end{figure} 
\begin{figure}[htbp]
	\centering
	\includegraphics[width=0.5\linewidth]{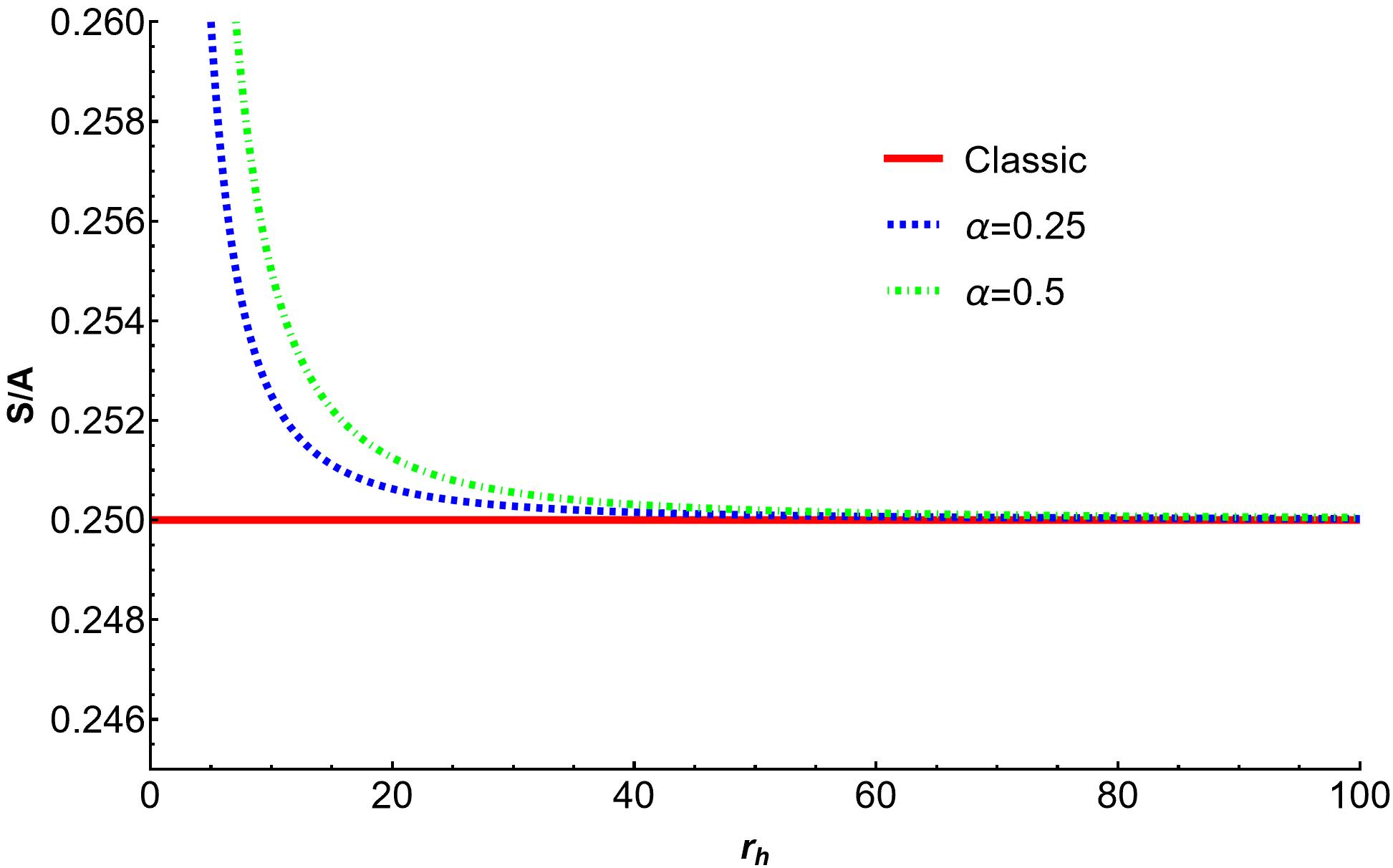}
	\caption{Entropy of quantum-corrected and Schwarzschild black hole in 6-dimensional space.}
	\label{S6}
\end{figure}

The event horizon area for a $d$-dimensional sphere is given by $A = \Omega_d r_h^{d-1}$ \cite{fernandes2023grand}. Having defined the surface area of the $n$-dimensional unit sphere in the previous section, we can re-express the horizon radius $r_h$ by the area $A$ and substitute it into the entropy. We then plot the ratio of entropy to surface area, $S/A$, against $r_h$. In the classical limit, we expect this ratio to be a constant value $S/A = 1/4$ \cite{witten2025introduction}, representing the standard Bekenstein-Hawking area law. As illustrated in Figs.~\ref{S5} and \ref{S6}, the quantum-corrected entropy asymptotically approaches the classical result as the horizon radius $r_h$ increases in higher-dimensional cases. At the same time, we find that as the number of dimensions increases, the contribution of the quantum correction term gradually decreases. This indicates that in high-dimensional space, the entropy of a black hole converges to the classical case, which is the dominant 1/4 term.

\subsection{Heat Capacity}
\begin{figure}[htbp]
	\centering
	\includegraphics[width=0.5\linewidth]{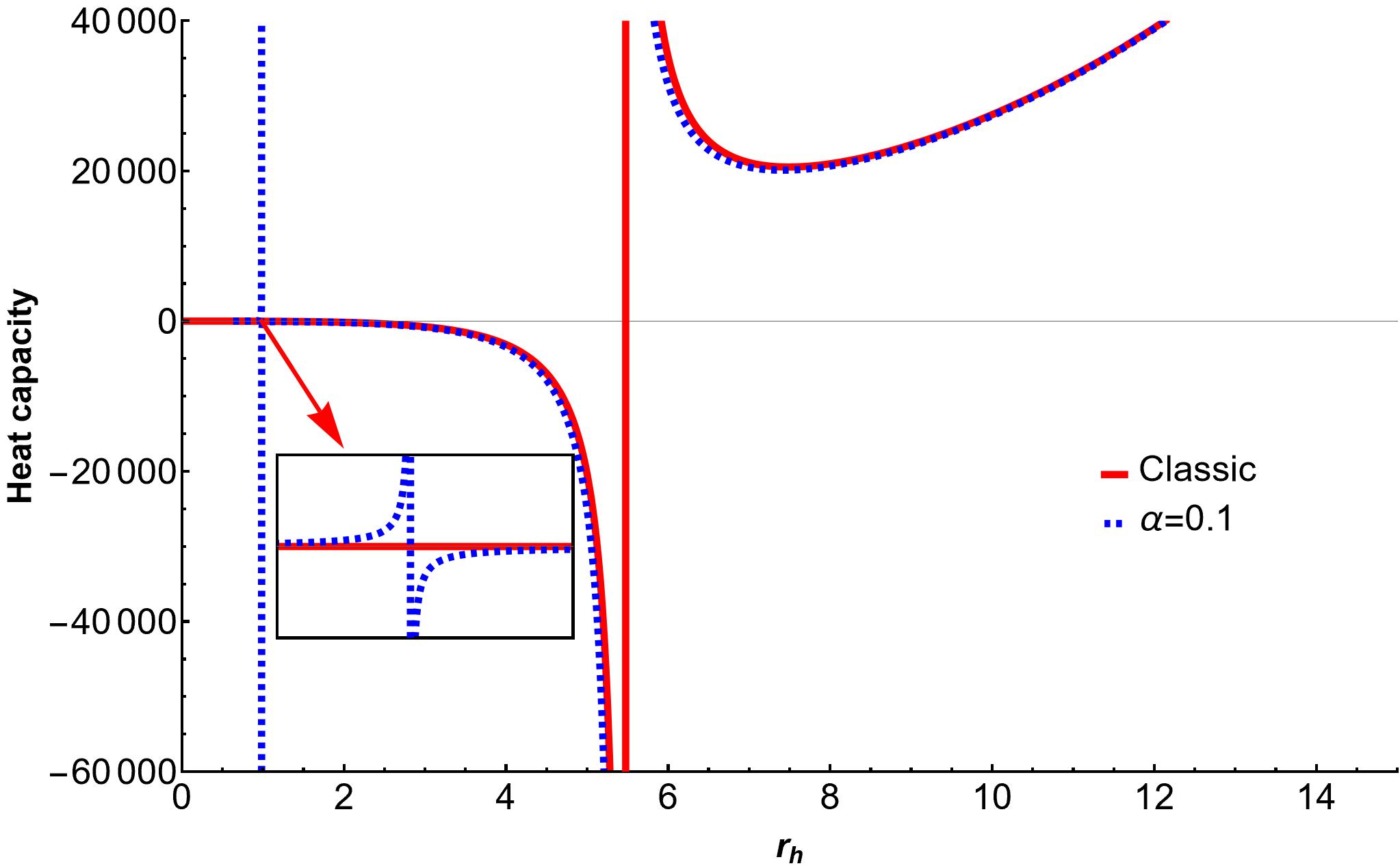}
	\caption{Heat capacity of the 5-dimensional quantum-corrected and Schwarzschild black hole.}
	\label{fig:C5}
\end{figure}

Having determined the entropy and temperature in the preceding sections, the heat capacity is calculated using the relation \cite{chatzifotis2023thermal}
\begin{equation}
C=T\left(\frac{\partial S}{\partial r_{h}}\right)\bigg/\left(\frac{\partial T}{\partial r_{h}}\right).
\label{Heat capacity}
\end{equation}
To recover the classical result, we take the limit where the correction parameter $\alpha$ vanishes. As an illustrative example, the heat capacity in five-dimensional spacetime is given by
\begin{equation}
C_5 = \frac{3\pi^{2}r_{h}^{4} \left[ 3r_{h}^{2} - 9\alpha + r_{h}\sqrt{r_{h}^{2}-4\alpha} (\alpha \Lambda-3) \right]}{2 \left[ 3r_{h}^{2}\sqrt{r_{h}^{2}-4\alpha} + 9\alpha \sqrt{r_{h}^{2}-4\alpha} + r_{h}^{3}(\alpha \Lambda-3) \right]}.
\label{C5}
\end{equation}
In Fig.~\ref{fig:C5}, we adopt the parameter values $\alpha=0.1$ and $\Lambda=-0.1$ to investigate the properties of the quantum-corrected black hole. When $r_{h}$ is relatively large, the quantum correction term approaches the classical result. However, when the radius is small, we observe a discontinuity in the quantum-corrected black hole. This feature indicates a thermodynamic phase transition in which the small black hole transforms from a thermally stable phase (positive heat capacity) to an unstable phase (negative heat capacity), a behavior absent in the classical black hole. 
The same phenomenon occurs for black holes with quantum corrections in other dimensions. We test \(5, 6\) and \(7\)-dimensional spacetime and find that the quantum correction term contributes one extra phase transition at small radii.

Regarding the phase transitions occurring at large radii for both classical and quantum-corrected heat capacities, we further investigate the dimensional dependence of these thermal behaviors. The corresponding results are summarized in Table~\ref{tab:horizon_radii}. It is observed that the position of the phase transition—characterized by the divergence of the heat capacity—shifts toward a larger $r_h$ as the spacetime dimension increases for both black hole models. Furthermore, the discrepancy between the classical and quantum-corrected capacities at large $r_h$ decreases in higher dimensions.
\begin{table}[htbp]
	\centering
	\caption{Discontinuity location of $r_h$ between classical and quantum-corrected (QC) black holes in different dimensions. The last column shows the discrepancy in \(r_h\) at the divergence point of heat capacity for classical versus quantum-corrected black holes. The quantum correction parameter $\alpha$ is fixed to 0.1}
	\label{tab:horizon_radii}
	\renewcommand{\arraystretch}{2.0} 
	\setlength{\tabcolsep}{12pt}
	
	\begin{tabular}{ccccc}
		\hline\hline
		\textbf{(d+1)D} & \textbf{Classical} & \textbf{QC (large $r_h$)} & \textbf{QC (small $r_h$)} & \textbf{Difference} \\
		\hline
		5 & 5.47723 & 5.40128 & 0.98353 & 0.07595 \\
		6 & 7.74597 & 7.70001 & 0.92323 & 0.04596 \\
		7 & 10.00000 & 9.96719 & 0.89356 & 0.03281 \\
		\hline\hline
	\end{tabular}
\end{table}

\begin{figure}
	\centering
	\includegraphics[width=0.5\linewidth]{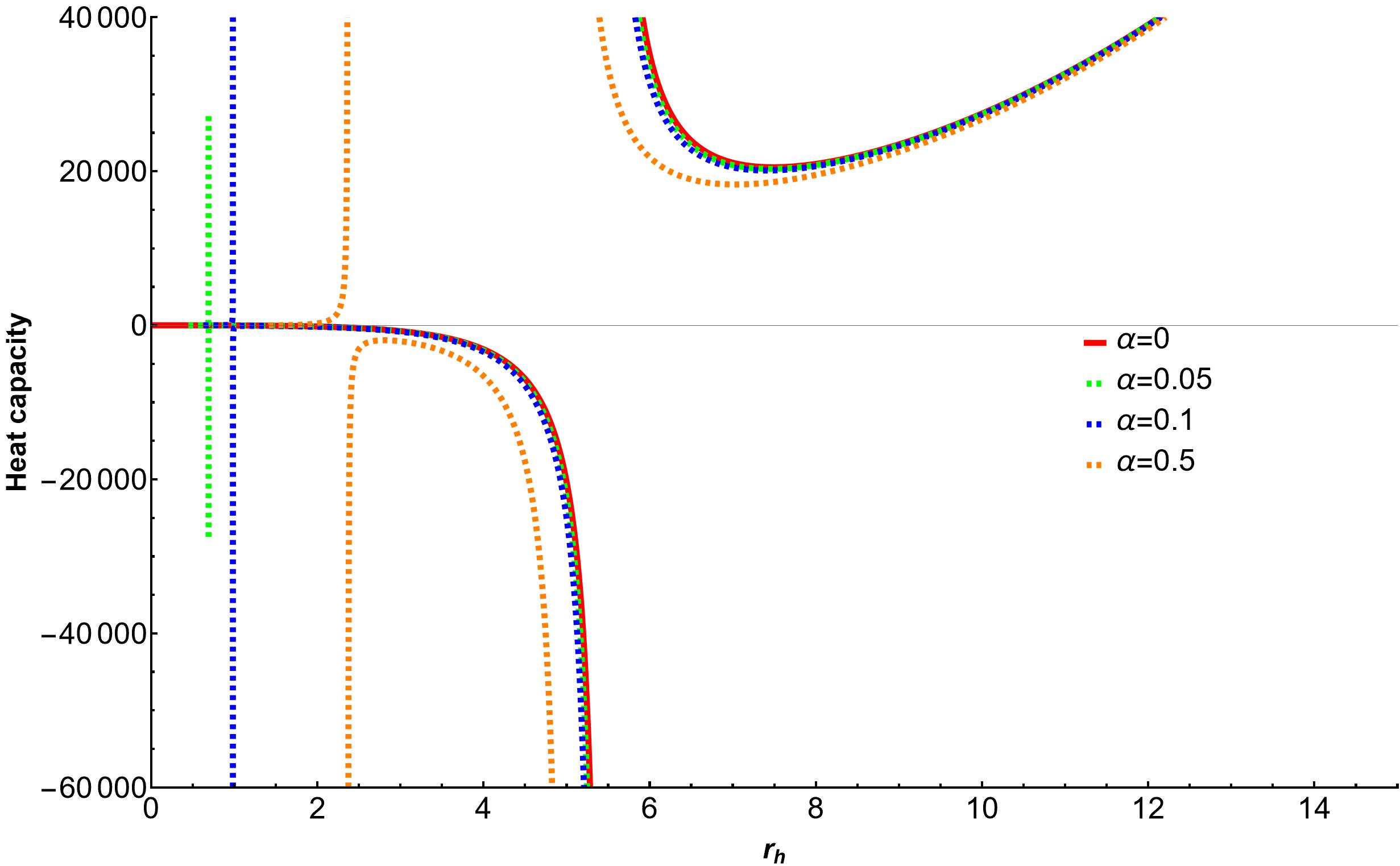}
	\caption{Heat capacity of the 5-dimensional quantum-corrected black hole with different values of $\alpha$ and Schwarzschild black hole.}
	\label{C53in1}
\end{figure}

In addition, we also compare the effect of different quantum correction parameters $\alpha$ in heat capacity (see Fig.~\ref{C53in1}). When the value of $\alpha$ increases, the position of the phase transition in the quantum-corrected heat capacity shifts toward smaller horizon radii. Consequently, as the correction parameter increases, the difference between the quantum-corrected heat capacity phase transition and the classical case becomes more significant. At the same time, for the additional phase transition occurring at smaller radii, an increase in $\alpha$ causes it to gradually approach the position of the phase transition at larger radii. We list the positions of phase transition under classical and quantum-corrected cases at \(5\) and \(6\)-dimensional spacetime, see Table~\ref{diffalpha} above.
\begin{table}[htbp]
	\centering
	\caption{Locations of phase transitions and their differences for different dimensions and $\alpha$ values.}
	\label{diffalpha}
	\renewcommand{\arraystretch}{1.5} 
	\setlength{\tabcolsep}{10pt} 
	\small 
	
	\begin{tabular}{cccccc}
		\hline\hline
		\textbf{(d+1)D} & \textbf{$\alpha$} & \textbf{Classical $r_h$} & \textbf{QC (large $r_h$)} & \textbf{QC (small $r_h$)} & \textbf{Difference} \\
		\hline
		\multirow{4}{*}{5} 
		& 0.00   & 5.47723 & NA      & NA      & NA \\
		& 0.05   & NA      & 5.44000 & 0.69061 & 0.03723 \\
		& 0.10   & NA      & 5.40128 & 0.98353 & 0.07595 \\
		& 0.50   & NA      & 5.00931 & 2.36853 & 0.46792 \\
		\cline{2-6} 
		\multirow{4}{*}{6}
		& 0.00   & 7.74597 & NA      & NA      & NA \\
		& 0.05   & NA      & 7.72318 & 0.65092 & 0.02279 \\
		& 0.10   & NA      & 7.70001 & 0.92323 & 0.04596 \\
		& 0.50   & NA      & 7.49819 & 2.11850 & 0.24778 \\
		\hline\hline
	\end{tabular}
\end{table}

The Gibbs free energy of the black hole is 
\begin{equation}
G=M-TS,
\label{Free energy}
\end{equation}
where $G$ can be expressed as a function of $r_{h}$. We combine the mass of black hole $M$, entropy from Eq. (\ref{entropy}) and Hawking temperature from Eq. (\ref{eq:Hawking_temperature}) and substitute them into Eq. (\ref{Free energy}). 

\begin{figure}[htbp]
    \centering
    \includegraphics[width=0.5\linewidth]{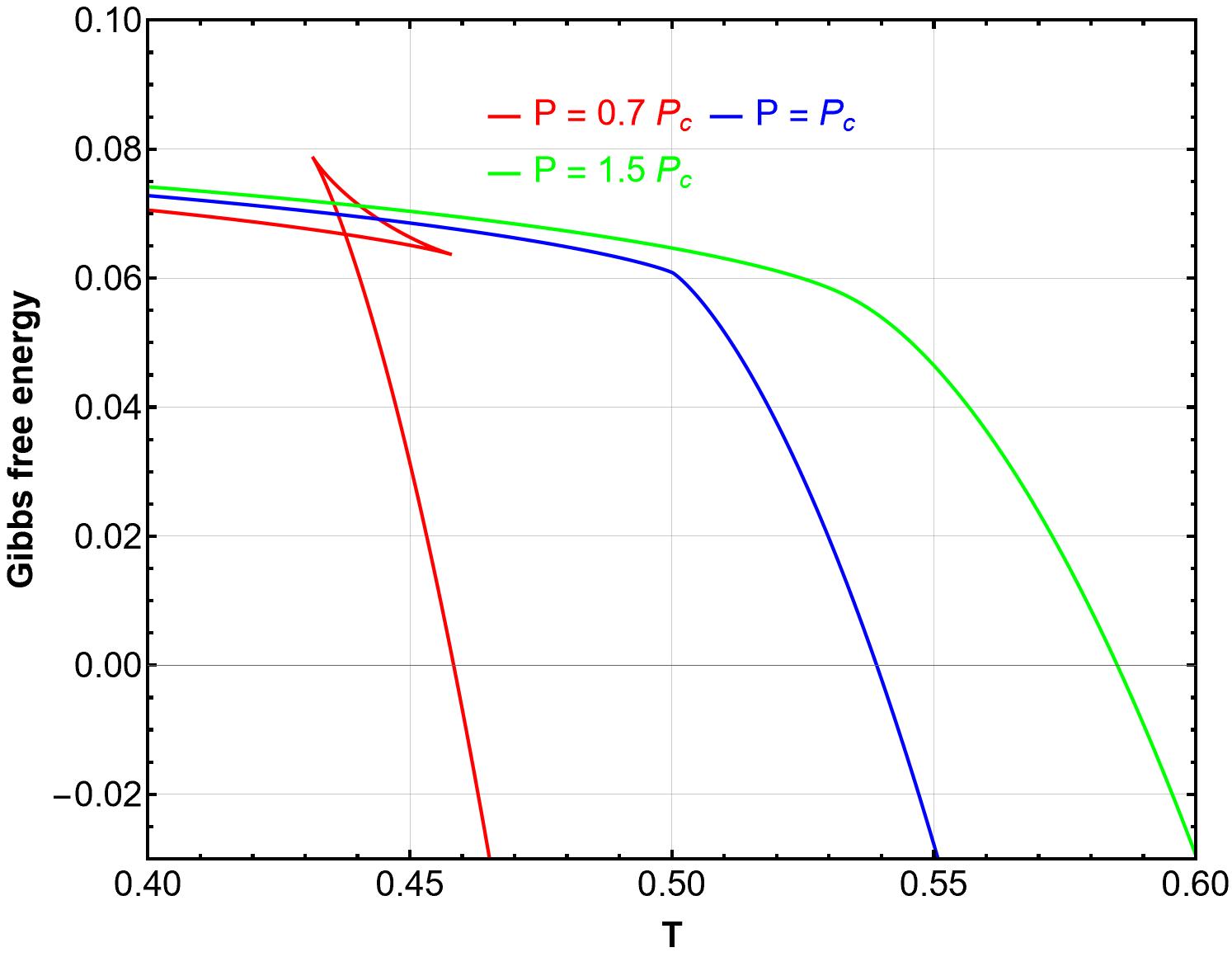}
    \caption{Gibbs free energy against temperature in 4-dimensional space}
    \label{G5}
\end{figure}

In Fig.~\ref{G5}, when $P<P_{c}$ (red curve), the image will exhibit a swallow tail behavior to indicate that there is a first order transition in the system. If we start increasing the temperature from $T = 0.40$, the system follows the small black hole branch (with lower Gibbs free energy) until it reaches the intersection with the large black hole branch. This intersection marks the first-order phase transition, where the system jumps from the small black hole phase to the thermodynamically favored large black hole phase with lower free energy. At the critical pressure $P=P_c$ (blue curve), the swallowtail structure vanishes, corresponding to a second-order phase transition at the critical point. For pressures above the critical value ($P>P_c$, green curve), the curve becomes monotonic, with no phase transition occurring and only a single black hole phase present across all temperatures. This behavior is qualitatively similar in higher-dimensional spacetimes.

\section{Critical Index}\label{Critical index}
Given the relation between pressure $P$ and the cosmological constant $\Lambda$, we adopt the identification $\Lambda = -8\pi P$. The thermodynamic volume in $(d+1)$-dimensional spacetime is given by $V = \Omega(d) r^d$, which allows us to express the horizon radius $r$ as a function of $V$. Substituting these two quantities into the Hawking temperature expression, we can derive the function $P(T,V)$. The critical points of temperature, volume and pressure $T_{c}$, $V_{c}$, $P_{c}$ are given by the following equations
\begin{equation}
\frac{\partial P}{\partial V}|_{T}=0,
\label{cpoint1}
\end{equation}
and 
\begin{equation}
\frac{\partial ^{2} P}{\partial V^{2}}|_{T}=0.
\label{cpoint2}
\end{equation}
The compressibility $Z$ and its value at the critical point\cite{10.1119/1.1986710} are given by
\begin{equation}
Z(V,T)=\frac{PV}{T},Z_{c}=\frac{P_{c}V_{c}}{T_{c}}.
\label{Z,Zc}
\end{equation}
$Z=1$ indicates that there is no interaction between the composing particles, and we find the critical compressibility $Z_{c}$ is related to the quantum-correction factor $\alpha$, see Table \ref{critical_values}.

\begin{table}[htbp]
	\centering
	\caption{Critical values $P_c$, $V_c$, $T_c$ and $Z_c$ for different dimensions.}
	\label{critical_values}
	\renewcommand{\arraystretch}{3} 
	\setlength{\tabcolsep}{10pt} 
	\small 
	
	\begin{tabular}{ccccc}
		\hline\hline
		\textbf{(d+1)-dimension} & \textbf{$P_c$} & \textbf{$V_c$} & \textbf{$T_c$} & \textbf{$Z_c$} \\
		\hline
		5 & $\dfrac{\alpha(49 - 18\sqrt{7})}{48\sqrt{7}\,\pi \alpha^{2}}$ 
		& $648\ \pi^{2} \alpha^{2}$ 
		& $\dfrac{\sqrt{2}}{9\pi\sqrt{\alpha}}$ 
		& $\dfrac{243\pi^{2}\sqrt{\alpha} \cdot \alpha(49-18\sqrt{7})}{2\sqrt{14}}$ \\
		6 & $\dfrac{\alpha(9 - 5\sqrt{3})}{8\sqrt{3}\,\pi \alpha^{2}}$ 
		& $\dfrac{8192}{3}\,\pi^{2} \alpha^{5/2}$ 
		& $\dfrac{1}{4\pi\sqrt{\alpha}}$ 
		& $\dfrac{4096\pi^{2}\alpha \cdot \alpha(9-5\sqrt{3})}{3\sqrt{3}}$ \\
		7 & $\dfrac{\sqrt{\frac{5}{33}}\;\alpha(121 - 9\sqrt{165})}{48\pi \alpha^{2}}$ 
		& $3375\,\pi^{3} \alpha^{3}$ 
		& $\dfrac{4}{3\sqrt{15}\,\pi\sqrt{\alpha}}$ 
		& $\dfrac{16875\pi^{3} \alpha^{3/2} \cdot \alpha(121-9\sqrt{165})}{64\sqrt{11}}$ \\
		\hline\hline
	\end{tabular}
\end{table}
The reduced variables are determined by
\begin{equation}
\tilde{t}=\frac{T-T_{c}}{T_{c}}, \tilde{v}=\frac{V-V_{c}}{V_{c}}, \tilde{p}=\frac{P}{P_{c}}.
\label{redvar}
\end{equation}
To investigate the behavior of quantities near the critical point, we introduce the critical component\cite{lin2024effective}
\begin{equation}
C_{V}=T (\frac{\partial S}{\partial T})\propto |\tilde{t}|^{-\alpha},
\label{cc1}
\end{equation}
\begin{equation}
\eta =V_{1}-V_{2}\propto |\tilde{t}|^{\beta},
\label{cc2}
\end{equation}
\begin{equation}
k_{T} =-\frac{1}{V}\frac{\partial V}{\partial P}\propto |\tilde{t}|^{-\lambda},
\label{cc3}
\end{equation}
\begin{equation}
|P-P_{c}|\propto |V-V_{c}|^{\chi}.
\label{cc4}
\end{equation}

In Sec.~\ref{Entropy}, we have shown that the entropy is independent of the temperature in all dimensions, leading to $\alpha=0$. In addition, if we substitute the reduced variables defined above into $P(T,V)$ and expand for small $\tilde{t}$ and $\tilde{v}$, the function of reduced pressure $\tilde{P}$ takes the form
\begin{equation}
\tilde{p}=1+(C_{1}-C_{2}\tilde{v})\tilde{t}-C_{3}\tilde{v}^{3}+\mathcal{O}(\tilde{v}^{4},\tilde{t}\tilde{v}^{2}),
\label{redp}
\end{equation}
the coefficients $C_{1}$, $C_{2}$, $C_{3}$ are constant, and if we take the derivative to $\tilde{v}$ at constant $\tilde{t}$, the equation becomes
\begin{equation}
d\tilde{p}=-(C_{2}\tilde{t}+3C_{3}\tilde{v}^{2})d\tilde{v}.
\label{redp2}
\end{equation}
Maxwell's law states that $\int VdP=0$, and pressure is constant at phase transition, so in this case
\begin{equation}
\tilde{p}(\tilde{t},\tilde{v_{1}})=\tilde{p}(\tilde{t},\tilde{v_{2}}), 
\label{redp3}
\end{equation}
\begin{equation}
\int_{\tilde v_{1}}^{\tilde v_{2}} (\tilde v +1)(\frac{\partial \tilde p}{\partial \tilde v} )_{\tilde t}d\tilde v=0.
\label{redp4}
\end{equation}
The expression of $\tilde{v_{1}}$ in terms of $\tilde{v_{2}}$ can be determined by calculating these two equations. We find a nontrivial solution and put this solution back into Eq. (\ref{redp3}) leads to
\begin{equation}
\tilde{v_{1}}=-\tilde{v_{2}}=\frac{C_{2}}{C_{3}}\sqrt{-\tilde{t}}.
\label{beta}
\end{equation}
Thus, the value of $\eta$ that we define in Eq. (\ref{cc2}) is proportional to $|\tilde{t}|^{\frac{1}{2}}$, which means $\beta=\frac{1}{2}$.
To find $k_{T}$, we need to change the parameters in (\ref{cc3}) into $\tilde{v}$ and $\tilde{p}$
\begin{equation}
\frac{\partial P}{\partial V} =\frac{\partial P}{\partial \tilde v} \frac{\partial \tilde v}{\partial V}=\frac{P_{c}}{V_{c}} \frac{\partial \tilde p}{\partial \tilde v},
\label{kTchange1}
\end{equation}
then $k_{T}$ becomes
\begin{equation}
k_{T}=-\frac{1}{\tilde v+1} \cdot \frac{1}{P_{c}\frac{\partial \tilde p}{\partial \tilde v} }.  
\label{kTchange2}
\end{equation}
We can also ignore the $\frac{1}{P_{c}}$ term in expression since it is a constant. After expanding $k_{T}$ in powers of $\tilde{v}$, we find that the expression is proportional to $\frac{1}{\tilde{t}}$, i.e. $\lambda=1$. Eq. (\ref{cc4}) can also be written as $|\tilde{p}-1|\propto |\tilde{v}|^{\chi}$ by combining with Eq. (\ref{redvar}) and removing the constant term. When we set $\tilde{t}=0$ in the critical isotherm condition, then
\begin{equation}
|\tilde{p}-1|\propto \tilde{v}^{3}, \chi=3.
\label{kai}
\end{equation}
To summarize, the critical components are $\alpha=0$, $\beta=\frac{1}{2}$, $\lambda=1$ and $\chi=3$ and their values are independent of the dimension.

\section{Conclusion}\label{Conclusion}
In this work, we have constructed the quantum-corrected Oppenheimer–Snyder black hole solution with a cosmological constant in $(d+1)$-dimensional spacetime, as shown in Eq.~\ref{dsfinal}. The interior spacetime describing the collapsing dust is governed by a $(d+1)$-dimensional loop quantum cosmology setup\textcolor{blue}{\cite{Li:2025bzl}}. The external region is assumed to be static and spherically symmetric. By imposing continuity of the metric and extrinsic curvature across the stellar surface via the Darmois–Israel junction conditions, the exterior metric is uniquely determined.

We calculated the thermodynamic properties of a black hole in an arbitrary \((d+1)\)-dimensional spacetime and observed the effect of the cosmological constant on these thermodynamic properties in the AdS spacetime. We find that the temperature of a black hole with quantum corrections rises to a peak when the event horizon radius is small, rather than diverging as in the classical case. The entropy with quantum corrections is independent of the cosmological constant in any dimension; it gradually approaches the classical limit as the radius $r_{h}$ increases.

For heat capacity, quantum corrections lead to an additional phase transition at smaller radii. We also calculated the difference between the specific heat capacities under quantum corrections and the classical case at the discontinuity points for larger radii as the dimension increases. We find that this difference gradually decreases with increasing dimension, indicating that the specific heat capacity under quantum corrections converges to the classical shape at larger radii in higher dimensions.


We calculated the critical exponents for various dimensions, which describe the behavior of physical quantities at the critical point. The critical exponents associated with the thermodynamic phase transition still satisfy the universal scaling laws,indicating that the critical behavior is insensitive to quantum corrections and spacetime dimension. In addition, when the pressure of Gibbs free energy is below the critical value, it can exhibit an extra phase transition from the small black hole to the large black hole phase.

Through the analysis of black hole thermodynamics, the Hawking temperature of small-radius black holes can tend to zero, implying the possible existence of stable black hole remnants. Recent studies on loop quantum gravity show that the covariant effective model can avoid spacetime singularities, and its derived Hawking radiation and greybody factors provide new insights into the final fate of black holes (e.g., black hole-white hole transition and stable remnants) \cite{Belfaqih:2025eak}. Future research can focus on calculating the Hawking radiation spectrum and greybody factors in our model, comparing with LQG-related results to clarify the evolutionary fate of quantum black holes.

\acknowledgments
This work is supported by National Natural Science Foundation of China (NSFC) with Grant No. 12275087.

\bibliography{ref}
\end{document}